\documentclass[aip,ajp,amsmath,amssymb,nofootinbib,preprint,nobibnotes]{revtex4-1}

\usepackage{graphicx}
\usepackage[utf8]{inputenc}
\usepackage[T1]{fontenc}
\usepackage{mathptmx}
\usepackage[colorlinks=true,linkcolor=blue,anchorcolor=blue,citecolor=blue,filecolor=blue,urlcolor=blue,breaklinks]{hyperref}
\usepackage{physics}

\makeatletter
\newcommand*{\fnsymbolss}[1]{%
	\ensuremath{%
	\ifcase#1 
	\or *\or \dagger\or \ddagger\or %
	\mathsection\or \mathparagraph\or \|\or **\or \dagger\dagger %
	\or \ddagger\ddagger \else\@ctrerr\fi}}
\makeatother

\newcommand*{\citen}{}
\DeclareRobustCommand*{\citen}[1]{
	\begingroup
		\romannumeral-`\x
		\setcitestyle{numbers}%
		\cite{#1}%
	\endgroup
}

\begin{document}

\title{Kramers-Kronig relations via Laplace formalism and $L^1$ integrability}

\author{Marco Prevedelli}
\affiliation{Department of Physics and Astronomy ``Augusto Righi'', University of Bologna, 40127 Bologna, Italy}
\author{Alessio Perinelli}
\affiliation{Department of Physics, University of Trento, 38123 Trento, Italy}
\affiliation{TIFPA-INFN, University of Trento, 38123 Trento, Italy}
\author{Leonardo Ricci}
\email[]{leonardo.ricci@unitn.it}
\affiliation{Department of Physics, University of Trento, 38123 Trento, Italy}

\date{\today}

\begin{abstract}
Kramers-Kronig relations link the real and imaginary part of the Fourier transform of a well-behaved causal transfer function describing a linear, time-invariant system. From the physical point of view, according to the Kramers-Kronig relations, absorption and dispersion become two sides of the same coin. Due to the simplicity of the assumptions underlying them, the relations are a cornerstone of physics. The rigorous mathematical proof was carried out by Titchmarsh in 1937 and just requires the transfer function to be square-integrable ($L^2$), or equivalently that the impulse response of the system at hand has a finite energy. Titchmarsh's proof is definitely not easy, thus leading to crucial steps that are often overlooked by instructors and, occasionally, prompting some authors to attempt shaky shortcuts. Here we share a rigorous mathematical proof that relies on the Laplace formalism and requires a slightly stronger assumption on the transfer function, namely its being Lebesgue-integrable ($L^1$). While the result is not as general as Titchmarsh's proof, its enhanced simplicity makes a deeper knowledge of the mathematical aspects of the Kramers-Kronig relations more accessible to the audience of physicists.
\end{abstract}

\maketitle

\section{Introduction}
Linear and time-invariant (LTI) systems are among the most basic, widespread and studied models in physics. Restricting the discussion to one-dimensional functions of time, an LTI system is characterized by a transfer function, $G(t)$, whose convolution with an input function provides the output function. In the frequency domain, this property assumes an even simpler form: in the Fourier or Laplace formalism, the transform of the output is given by the product of the transforms of the input and the transfer function.

As if that were not enough, requiring the transfer function $G(t)$ to be causal and have a finite energy, as it is always the case in the real physical world, produces a spectacular result~\cite{Nussenzveig1972}: the mutual dependency of the physical descriptions underlying absorption and dispersion, which describe how a system reacts to an input in terms of energy and phase delay, respectively. So the rainbow exists because at some wavelengths other than visible ones water is opaque, and vice versa. More specifically, the real and imaginary parts of the Fourier transform $\widetilde{G}_F(\omega)$ of $G(t)$, or susceptibility $\chi(\omega)$, are the Hilbert transforms of one another. The result was first derived, independently, by R.\ de L.\ Kronig~\cite{Kronig1926} and H.\ A.\ Kramers~\cite{Kramers1927}. However, the eponymous relations got a solid and definitive mathematical justification only with the work by E.\ C.\ Titchmarsh~\cite{Titchmarsh1948}, who proved the Kramers-Kronig relations to hold if and only if the causal transfer function $G(t)$ is square-integrable; i.e.\ it belongs to $L^2$:
\begin{equation}
	\label{eq:L2}
	\int_0^\infty \left| G(t) \right|^2 \, \dd{t} < +\infty \,.
\end{equation}

The Kramers-Kronig relations, sometimes referred to as ``dispersion relations''~\cite{Holbrow1964,Bohren2010,YuffaScales2012,Dethe2019}, are a cornerstone of physics, whose implications are broadly investigated in a wide range of fields~\cite{Libbrecht2006,Horsley2017,Gulgowski2021,Stefanski2022} and thus go beyond the prototypical problem of interpreting the refraction index $n(\omega)$, for which they were first devised~\cite{Kronig1926}. Although Titchmarsh's contribution was recognized long ago~\cite{Toll1956}, a few decades after the formulation and proof of the relations, the awareness among the community of physicists about Titchmarsh's achievement was not unanimous. So Sharnoff in 1964 still argued~\cite{Sharnoff1964}: \emph{``It is paradoxical that although the Kramers-Kronig relations are so widely used, the literature contains neither a convincing proof of their general validity nor a careful discussion of sets of conditions under which they might be expected to hold.''}

The likely reason is that the Titchmarsh theorem in Fourier analysis--as the theorem is named---is definitely not straightforward to prove. Indeed most textbooks and papers citing it arrive just short of a complete proof when they typically take for granted the crucial and most difficult step: how to prove that the Fourier transform $\widetilde{G}_F(\omega)$, where the usually real frequency $\omega$ is extended to the complex plane, is analytic not only on the open, upper half-plane ${\text{Im}(\omega) > 0}$, but also on the real axis ${\text{Im}(\omega) = 0}$. So, for example, in J.\ D.\ Jackson's classic book on electrodynamics~\cite{Jackson1999} the step is justified with the words \emph{``On the real axis it is necessary to invoke the `physically reasonable' requirement that ${G(\tau)\to 0}$ as ${\tau \to \infty}$ to assure that ${\varepsilon(\omega)/\varepsilon_0}$ is also analytic there.''}\footnote{In Jackson's book, ${\varepsilon(\omega)/\varepsilon_0 - 1}$ corresponds to the Fourier transform of $G(\tau)$. The requirement ${G(\tau)\to 0}$ as ${\tau \to \infty}$ is physically reasonable because dissipative mechanisms loom everywhere, so the impulse response of any real system must fade out some time. While this behavior translates, as a consequence of Parseval's theorem, in $\widetilde{G}_F(\omega)$ being vanishing too as ${|\omega| \to \infty}$, how this implies the analyticity on the real axis is less immediate.} A similar argument is used by Bechhoefer~\cite{Bechhoefer2011} who, in order to provide \emph{``a brief derivation of the Kramers-Kronig relations''}, states: \emph{``for simplicity, we will also assume that $G(\omega)$ has no poles on the real axis''}. Finally, the likewise classic book by Landau and Lifshitz on Statistical Physics~\cite{LandauLifshitz1980} proposes a proof that is based on a previous theorem, proved by N. N. Me\u{\i}man, which derives asymptotic properties of $\chi(\omega)$ as a consequence of Cauchy's argument principle~\cite{BrownChurchill2009}. However, exactly as in the previous cases, also this theorem implicitely takes for granted the analyticity of $\chi(\omega)$ on the real axis ${\text{Im}(\omega) = 0}$.

The arduousness of the proof, combined with the importance of the result, has prompted several attempts to find simpler approaches. Searching the internet is likely to provide alleged solutions from non-peer-reviewed sources and, occasionally, peer-reviewed ones~\cite{Hu1989}, which invariably fail to live up to the promises. A common trait of these attempts is their being based on the combination of two ingredients: the convolution theorem, and the Fourier transform of the Heaviside step function $\theta(t)$, in fact the very expression of causality. The convolution theorem states that the Fourier transform of the convolution of two functions or distributions of time is equal to the product of the Fourier transforms of the factors. Due to the symmetry of the Fourier transform and its inverse, the theorem can be read the other way round, i.e.\ the inverse Fourier transform of the convolution of two functions or distributions of frequency is equal to the product of the inverse Fourier transforms of the factors multiplied by ${2 \pi}$. With regard to the Fourier transform of the Heaviside step function $\theta(t)$, $\chi(\omega)$ is given by the sum of the two distributions ${\pi \delta(\omega)}$ and ${i\,\text{P}\,\omega^{-1}}$, where $\text{P}$ indicates the Cauchy principal value. Consequently, starting from the expression ${G(t) = \theta(t)\,G(t)}$, which is true because of causality, and applying the ``inverse'' version of the convolution theorem one can derive the expression
\begin{equation}
	\chi(\omega) \star \text{P}\frac{1}{\omega} = -i \pi \chi(\omega) \,,
\end{equation}
which corresponds to the Kramers-Kronig relations being written as a convolution.

This result is well known (see, for example, Sec. 1.8 of Ref.\citen{Nussenzveig1972}), but it definitely not easy to derive, the most difficult part being the conditions under which the convolution theorem holds. In fact, the derivation of the expression above requires the theory of distributions, and it is far from being more elementary than Titchmarsh's one. On the other hand, the alleged solutions mentioned above go straight to the final expression, disregarding essential aspects of validity, which essentially coincide with those set by the Titchmarsh theorem and whose omission leads to miscalculations. For example, referring to the attempt by Hu, described in the 1989 paper \emph{``Kramers-Kronig (relations) in two lines''}~\cite{Hu1989}, one could try to verify whether the relations work when the function $\hat{Y}(t)$ is given by a constant value, or by the sign function ${\text{sgn}(t) = 2\cdot\theta(t) - 1}$. They do not, as a direct evaluation promptly shows. To conclude, the Kramers-Kronig relations are a powerful tool that stems from linearity, causality, and energy boundedness: proving the link is arduous and admits no shortcuts.

Here we show that the Kramers-Kronig relations can alternatively be derived in a way that is simpler than Titchmarsh's one. The derivation relies on the intrinsically causal Laplace formalism and on the assumption that the transfer function is Lebesgue-integrable rather than square-integrable, i.e.\ belonging to $L^1$ (rather than $L^2$):
\begin{equation}
	\label{eq:L1}
	\int_0^\infty \left| G(t) \right| \, \dd{t} < +\infty \,.
\end{equation}
However, simplicity comes at a cost: as it will be shown later, ${G(t) \in L^1}$ provides a sufficient condition for the Kramers-Kronig relations to hold, rather than a necessary and sufficient one as in Titchmarsh's formulation. Therefore, though providing a solid path to the Kramers-Kronig relations, the present proof does not replace Titchmarsh's one, which remains unsurpassed.

In the following, after a brief review of the Laplace and Fourier transforms and the related properties that are functional to the proof, we introduce Laplace-transformable, Lebesgue-integrable functions, for which the Kramers-Kronig relations are thereupon proved. The limit of the present proof compared with Titchmarsh's classic one is finally discussed.

\section{A review of Laplace and Fourier transforms}
\label{sec:laplace_transform}
To aid the reader, and despite their being common knowledge, we summarize here the definition and some properties of the Laplace transform that are important for the discussion below: analyticity, Riemann-Lebesgue lemma, and the inversion of the transform via Bromwich, or Fourier-Mellin, integral. For the same purpose, at the end of the section we also review the definition of the Fourier transform and when a function is Fourier-transformable.

Let $s$ be a complex variable and $s^\prime$, $s^{\prime\prime}$ its real and imaginary part, respectively. By definition, a function $f(t)$ is causal if, for all ${t < 0}$, ${f(t) = 0}$, and is locally integrable if the integration on any compact subset\footnote{For a function of a real variable, compact is equivalent to closed and bounded.} of its domain is finite. The Laplace transform $\widetilde{F}_L(s)$ of a causal and locally integrable $f(t)$ is defined as
\begin{equation}
	\label{eq:Laplace_transform_definition}
	\widetilde{F}_L(s) \equiv \int_0^\infty e^{-st} f(t) \, \dd{t} \,.
\end{equation}
The complex half-plane where the above integral absolutely converges (\emph{\'{a} la} Lebesgue), i.e.\ the set of complex numbers $s$ such that
\begin{equation}
	\label{eq:Laplace_transform_absolute_convergence}
	\int_0^\infty \left| e^{-st} \, f(t) \right| \, \dd{t} = \int_0^\infty e^{-s^\prime t} \, \left| f(t) \right| \, \dd{t} < \infty \,,
\end{equation}
is left-bounded by the so-called abscissa of absolute convergence $\lambda_0$: $\lambda_0$ is the minimum real number such that absolute convergence occurs for any ${s^\prime > \lambda_0}$. A function $f(t)$ that is causal, locally integrable, and has a finite $\lambda_0$ is henceforth referred to as a Laplace-transformable function.

The main consequence of the absolute convergence condition is the analyticity of the Laplace transform $\widetilde{F}_L(s)$ in the half-plane of absolute convergence, i.e.\ for ${s^\prime > \lambda_0}$. Another consequence is the Riemann--Lebesgue lemma, which follows from Lebesgue's dominated convergence theorem:
\begin{equation}
	\label{eq:RiemannLebesguelemma}
	\lim_{s^\prime \to +\infty} \widetilde{F}_L(s) = 0 \,.
\end{equation}

Finally it is worth stating the general expression for the inverse Laplace transform, which corresponds to the Bromwich, or Fourier-Mellin, integral:
\begin{equation}
	\label{eq:Laplace_inverse_transform}
	f(t) = \frac{1}{2\pi i} \int_{a-i\infty}^{a +i \infty} \widetilde{F}_L(s) e^{st} \, \dd{s} \,,
\end{equation}
where $a$ is any constant real number such that $a > \lambda_0$.

The Fourier transform of a function $f(t)$, not necessarily causal, is
\begin{equation}
	\label{eq:Fourier_transform_definition}
	\widetilde{F}_F(\omega) \equiv \int_{-\infty}^\infty e^{i \omega t} f(t) \, \dd{t} \,,
\end{equation}
where $\omega$ is a real frequency. Here ``the physicists's notation'' for the phasors of positive frequency, namely ${e^{-i\omega t}}$, is used.\footnote{The expression above carries out a ``projection'' of the original function $f(t)$ onto the phasor corresponding to the frequency $\omega$. In quantum mechanics, the projection of a wave-function onto another wave-function is a scalar product that requires the complex conjugation of the latter. This is the reason why, within the Fourier integral, the complex conjugate of the phasor appears.} Engineers typically use ${e^{j\omega t}}$ instead. The two notations are completely equivalent, due to the invariance of the real world under complex conjugation and the freedom we have in choosing the ``reference'' root of ${x^2 = -1}$: there are indeed two, $i$ and $-i$, or, better, $i$ and ${j = -i}$. As a result, one can recover the engineers' notation by replacing here and henceforth $i$ with $-j$ and, in particular, setting ${s = j \omega}$.

It is now worth mentioning that a common mistake consists of assuming square-integrability (${f(t) \in L^2}$). Indeed, $f(t)$ is Fourier-transformable if it is Lebesgue-integrable (${f(t) \in L^1}$), i.e.\ if it satisfies Eq.~(\ref{eq:L1}) (with $f$ instead of $G$). Remarkably, ${f(t) \in L^2}$ does not necessarily imply ${f(t) \in L^1}$, and thus the Fourier-transformability of $f(t)$. An example is provided by ${f(t) = \theta(t-1)/t}$, which belongs to $L^2$ though not to $L^1$. Conversely, ${f(t) \in L^1}$ does not imply ${f(t) \in L^2}$ either: the function ${f(t) = \theta(t) \, \theta(1-t) /\sqrt{t}}$ provides a counterexample. On the other hand, it is well known that, in the case of square-integrability, Parseval's theorem holds and the inverse Fourier transform is essentially the same operator as the direct Fourier transform. The conundrum of a function $f(t)$ that belongs to $L^2$ but not to $L^1$ can be overcome by redefining the Fourier transform as follows~\cite{Rudin1986}. One can consider the sequence of functions
\begin{equation}
	f_n(t) = f(t) \left[\theta(t+n) - \theta(t-n) \right] \,,
\end{equation}
where $n$ is a positive integer number. Each function $f_n(t)$, which can be shown to belong simultaneously to $L^1$ and $L^2$, tends to $f(t)$ as ${n \to \infty}$ with respect to the $L^2$-norm given by ${\lVert f\rVert = \int_\mathbb{R}|f(t)|^2\,\dd{t}}$. Defining the Fourier transform of $\widetilde{F}_F(\omega)$ of $f(t)$ as the limit of the sequence of Fourier transforms $\widetilde{F}_{F,n}(\omega)$ when ${n \to \infty}$ eventually settles the problem.

\section{Laplace-transformable, Lebesgue-integrable functions}
\label{sec:laplace_transformable_L1_functions}
In the following discussion, besides being Laplace-transformable, the function $f(t)$ is assumed to be Lebesgue-integrable, i.e.\ to belong to $L^1$ and thus to satisfy
\begin{equation}
	\int_0^\infty \left| f(t) \right| \, \dd{t} < \infty \,.
\end{equation}
Comparing this last equation with Eq.~(\ref{eq:Laplace_transform_absolute_convergence}) requires the abscissa of absolute convergence $\lambda_0$ to be negative, so $\widetilde{F}_L(s)$ is analytic on the closed right-half plane (RHP), namely the set of $s$ such that ${s^\prime \geqslant 0}$.

Due to ${\lambda_0 < 0}$, one can set ${a=0}$ in Eq.~(\ref{eq:Laplace_inverse_transform}), so the integration occurs on the imaginary axis. The substitution ${s = -i\omega}$, where $\omega$ is a real variable, yields
\begin{equation}
	f(t) = \frac{1}{2\pi} \int_{-\infty}^{\infty} \widetilde{F}_L(-i\omega) e^{-i\omega t} \, \dd{\omega} \,.
\end{equation}

Upon noting that ${f(t) \in L^1}$ is the basic condition for the Fourier-transformability of $f(t)$, it is straightforward to recognize that $\widetilde{F}_L(-i\omega)$ is equal to the Fourier transform $\widetilde{F}_F(\omega)$ of $f(t)$:
\begin{equation}
	\label{eq:laplace_fourier_relation}
	\widetilde{F}_F(\omega) = \widetilde{F}_L(-i\omega) \,.
\end{equation}
The inverse is true as well: $f(t)$ being causal and Fourier--transformable implies its Fourier transform ${\widetilde{F}_F(\omega)}$ to correspond to the Laplace transform ${\widetilde{F}_L(s)}$, with $s^\prime = 0$, ${s^{\prime\prime} = -i\omega}$, and to this Laplace transform having ${\lambda_0 < 0}$.

\section{Kramers-Kronig relations}
\label{sec:kramers_kronig_relations}
\begin{figure}[h!]
	\centering
	\includegraphics[scale=1.2]{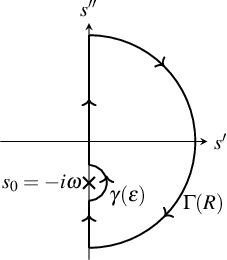}
	\caption{Integration path for the derivation of the Kramers--Kronig relations from $\widetilde{G}_L(s)/(s-s_0)$. The closed path is formed by two semicircular paths $\Gamma(R)$ and $\gamma(\varepsilon)$ of radii $R$ and $\varepsilon$, respectively, connected by two segments belonging to the imaginary axis. The path excludes the pole in $-i\omega$.}
	\label{fig:integration_path_kramers_kronig}
\end{figure}

We now suppose that, besides being causal, the transfer function $G(t)$ of an LTI system is Lebesgue-integrable. Upon setting a complex number ${s_0 = -i\omega}$ lying on the imaginary axis, we then consider the following function
\begin{equation}
	H(s,\omega) \equiv \frac{\widetilde{G}_L(s)}{s-s_0} = \frac{\widetilde{G}_L(s)}{s+i\omega} \,.
\end{equation}
Due to $\widetilde{G}_L(s)$ being analytic on the closed RHP, $H(s,\omega)$ is analytic on the closed RHP except at the point ${s = s_0 = -i\omega}$. By virtue of Cauchy residue theorem, an integration along the closed path shown in Fig.~\ref{fig:integration_path_kramers_kronig} yields a vanishing result because no poles lie within the path:
\begin{equation}
	\label{eq:path_integral}
	\int_{-iR}^{-i\omega-i\varepsilon} H(s,\omega) \, \dd{s} + \int_{\gamma(\varepsilon)} H(s,\omega) \, \dd{s} + \int_{-i\omega+i\varepsilon}^{iR} H(s,\omega) \, \dd{s} + \int_{\Gamma(R)} H(s,\omega) \, \dd{s} = 0 \,,
\end{equation}
where the $\Gamma(R)$, $\gamma(\varepsilon)$ are two semicircular paths of radii $R$ and $\varepsilon$, respectively, that are connected by the two linear segments joining the points on the imaginary axis of ordinates $-iR$, $-i\omega-i\varepsilon$, and $-i\omega+i\varepsilon$, $iR$.

Once ${\varepsilon \to 0^+}$ and ${R \to \infty}$, the sum of the integrals along the linear segments can be expressed as the Cauchy principal value of a single integral:
\begin{equation}
	\lim \limits_{\substack{\varepsilon \to 0^+\\R \to \infty}} \left(\int_{-iR}^{-i\omega-i\varepsilon} H(s,\omega) \, \dd{s} + \int_{-i\omega+i\varepsilon}^{iR} H(s,\omega) \, \dd{s}\right) =
	P \int_{-i\infty}^{+i\infty} H(s,\omega) \, \dd{s} = P \int_{-i\infty}^{+i\infty} \frac{\widetilde{G}_L(s)}{s + i\omega} \, \dd{s} \,.
\end{equation}
The integral on $\gamma(\varepsilon)$, which runs counterclockwise, is equal to $i\pi \widetilde{G}_L(-i\omega)$.

Now comes the crucial part of the theorem, namely to show that the integral along $\Gamma(R)$ vanishes as ${R \to \infty}$. Upon writing $s$ in polar coordinates as ${s = R\,e^{i\theta}}$, the integral can be written as
\begin{equation}
	\lim_{R \to \infty} \int_{\Gamma(R)} H(s,\omega) \, \dd{s} = \lim_{R \to \infty} \int_{\pi/2}^{-\pi/2} \frac{\widetilde{G}_L\left(R\,e^{i\theta}\right)}{R\,e^{i\theta} + i\omega} i\,R\,e^{i\theta} \, \dd{\theta} = \lim_{R \to \infty} \int_{\pi/2}^{-\pi/2} i\,\frac{s\,\widetilde{G}_L(s)}{s + i\omega} \, \dd{\theta} \,.
\end{equation}
For any $\theta$ within the interval ${\left(-\pi/2,\,\pi/2\right)}$ the limit ${R \to \infty}$ implies ${s^\prime \to +\infty}$, so ${\widetilde{G}_L(s)}$ tends to zero as a consequence of the Riemann--Lebesgue lemma expressed in Eq.~(\ref{eq:RiemannLebesguelemma}). Therefore, because given a real number ${\ell > 1}$, one has ${\left| s/(s+i\omega) \right| \leqslant \ell}$ as soon as ${R \geqslant \ell |\omega| / (\ell - 1)}$, the whole integral vanishes as well.\footnote{By writing ${s = R e^{i\theta}}$, one has
\begin{equation*}
	\left| \frac{s}{s+i\omega} \right| \leqslant \ell \Leftrightarrow \frac{\omega^2}{R^2} + 2 \frac{\omega}{R} \sin(\theta) + 1 \geqslant \frac{1}{\ell^2} \,.
\end{equation*}
Because, for any real number $a$, ${a \sin(\theta) \geqslant -|a|}$, it holds
\begin{equation*}
	\frac{\omega^2}{R^2} - 2 \frac{|\omega|}{R} + 1 \geqslant \frac{1}{\ell^2}\Leftrightarrow \frac{|\omega|}{R} \leqslant 1 - \frac{1}{\ell} \,.
\end{equation*}}

The argument used here is similar to Jordan's lemma, which states that if the maximum value of $\widetilde{G}_L(s)$ satisfies the Riemann--Lebesgue lemma, then, for ${t < 0}$, one has
\begin{equation}
	\lim_{R \to \infty} \int_{\Gamma(R)} \widetilde{G}_L(s) \, e^{st} \, \dd{s} = 0 \,.
\end{equation}
The main difference between Jordan's lemma and the present argument is therefore the factor $e^{st}$, which is here replaced with ${1/(s+i\omega)}$. In addition, while in Jordan's lemma the sign of the parameter $t$ plays a crucial role to achieve the convergence to zero of the integral, here the parameter $\omega$ plays no role.

Setting ${s = -i\nu}$, ${\nu \in \mathbb{R}}$, and remembering the relation between Laplace and Fourier transform expressed by Eq.~(\ref{eq:laplace_fourier_relation}) above, the path integral of Eq.~(\ref{eq:path_integral}) can then be rewritten as
\begin{equation}
	\widetilde{G}_F(\omega)=\frac{1}{i \pi} P\int_{-\infty}^{\infty}\frac{\widetilde{G}_F(\nu)}{\nu-\omega} \, \dd{\nu} \,.
\end{equation}

Separating the real and imaginary part of the Fourier transform by writing ${\widetilde{G}_F(\omega)=\widetilde{G}_F^\prime(\omega)+i\widetilde{G}_F^{\prime\prime}(\omega)}$ yields
\begin{equation}
	\begin{aligned}
		\widetilde{G}_F^\prime(\omega) &= \frac{1}{\pi}P\int_{-\infty}^{\infty}\frac{\widetilde{G}_F^{\prime\prime}(\nu)}{\nu-\omega} \, \dd{\nu},\\
		\widetilde{G}_F^{\prime\prime}(\omega) &= -\frac{1}{\pi}P\int_{-\infty}^{\infty}\frac{\widetilde{G}_F^\prime(\nu)}{\nu-\omega} \, \dd{\nu},\\
	\end{aligned}
\end{equation}
that corresponds to the well known Kramers--Kronig relations~\cite{Toll1956}, i.e.\ to $\widetilde{G}_F^\prime(\omega)$, $\widetilde{G}_F^{\prime\prime}(\omega)$ being the Hilbert transforms of one another.

\section{Discussion}
We mentioned above that the Kramers-Kronig relations were proven by Titchmarsh to be valid for any $L^2$, causal function $G(t)$. One might argue that the present proof of the Kramers-Kronig relations, in which the $L^2$ assumption is replaced with the $L^1$ assumption (see diagram below in Fig.~\ref{fig:proof_scheme}), is, due to its enhanced simplicity, superior to Titchmarsh's approach. Indeed, there are two reasons why the Titchmarsh theorem still makes up the unsurpassed way to achieve the Kramers-Kronig relations.

\begin{figure}[h!]
	\centering
	\includegraphics[scale=1.2]{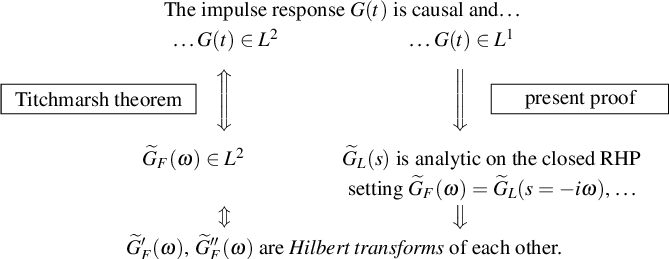}
	\caption{Diagram of the Titchmarsh's proof (left) and the present one (right).}
	\label{fig:proof_scheme}
\end{figure}

First, any function $\widetilde{G}_L(s)$ that is analytic on the closed RHP does not necessarily correspond to a causal, $L^1$ transfer function: as a major counterexample, a constant $\widetilde{G}_L(s)$ cannot be the Laplace transform of any regular function because it would violate the Riemann--Lebesgue lemma (a constant $\widetilde{G}_L(s)$ is, indeed, the Laplace transform of a distribution, namely a Dirac delta in the origin). For this reason, our approach to the Kramers-Kronig relations is one-way only. Conversely, the Titchmarsh theorem can be read also backwards: a function $\widetilde{G}_F(\omega)$ belonging to $L^2$ and whose real and imaginary parts are Hilbert transforms of one another does correspond to a causal, $L^2$ transfer function ${G(t) = \mathcal{F}^{-1}\left[\widetilde{G}_F(\omega)\right]}$. Second, proving the Kramers-Kronig relations for $L^2$ functions makes \emph{the use} of the theorem more handy, and this is what matters in physical applications.

However, as mentioned above, our approach provides an easier proof in all the cases in which the causal transfer function belongs to ${L^1 \cap L^2}$.

\section*{Conflict of interest}
The authors have no conflicts to disclose.

\bibliography{kramers_kronig_L1_proof}

\end{document}